\documentclass[12pt,a4paper]{article}
\usepackage{jheppub}
\usepackage{subfigure}
\usepackage{amsmath}

\newcommand{\be}{\begin{equation}}
\newcommand{\ee}{\end{equation}}
\newcommand{\beq}{\begin{equation}}
\newcommand{\eeq}{\end{equation}}
\newcommand{\ba}{\begin{eqnarray}}
\newcommand{\ea}{\end{eqnarray}}
\newcommand{\bea}{\begin{eqnarray}}
\newcommand{\eea}{\end{eqnarray}}
\newcommand{\nn}{\nonumber}
\newcommand{\ren}{R\'enyi\ }

\newcommand{\mt}[1]{\textrm{\tiny #1}}
\newcommand{\reef}[1]{(\ref{#1})}

\newcommand{\eg}{{\it e.g.,}\ }
\newcommand{\ie}{{\it i.e.,}\ }
\newcommand{\lp}{\ell_{\mt P}}

\newcommand{\labell}[1]{\label{#1}}

\arxivnumber{1305.nnnn}

\title{Holographic R\'enyi Entropies at Finite Coupling}
\author[a,b]{Dami\'an A. Galante}
\author[a]{and Robert C. Myers}

\affiliation[a]{Perimeter Institute for Theoretical Physics,\\
                                          Waterloo, Ontario N2L 2Y5, Canada}
\affiliation[b]{Department of Applied Mathematics, University of Western Ontario,\\
           London, Ontario N6A 5B7, Canada}

\emailAdd{dgalante@perimeterinstitute.ca}
\emailAdd{rmyers@perimeterinstitute.ca}

\abstract{
We compute R\'enyi entropies for a spherical entangling surface in
four-dimensional ${\cal {N}} = 4$ super-Yang-Mills at strong coupling using the
AdS/CFT correspondence. Incorporating the effects of the leading $\alpha'$
corrections to the low energy effective action of type IIB string theory, we
calculate the leading corrections in inverse powers of the 't Hooft
coupling (and the number of colours). The results are compared with known weak coupling
calculations. Setting the order of the \ren entropy $q$ to one, it reduces to the
entanglement entropy and the strong and weak coupling results match without any
corrections, as expected. In the limit of $q \rightarrow 0$, the relation
between the strong and weak coupling entropies is connected to the known
corrections for the thermal free energy in flat space. We also compute the
correction to the scaling dimension of twist operators.}

\begin{document}
\maketitle

\section{Introduction}

Entanglement is emerging as a fundamental phenomena in a wide variety of areas
ranging from quantum information and condensed matter physics, \eg
\cite{wenx,cardy0} to quantum gravity and string theory, \eg
\cite{sorkin,suss,rt1}. A useful probe for entanglement in quantum systems is
the entanglement entropy (EE). Given a subsystem $A$, one constructs the
reduced density matrix $\rho_A$ by tracing over the degrees of freedom in the
complement to $A$ for some global state or density matrix describing the full
system. Then, the EE is simply defined as the von Neumann entropy of
$\rho_\mt{A}$: $S_\mt{EE} \equiv - $tr$ (\rho_\mt{A} \log \rho_\mt{A})$. This
can be extended to a more general family of entanglement measures known as
entanglement R\'enyi entropies (ERE) \cite{renyi0,renyi1}. These are defined as
\begin{eqnarray}
S_q=\frac{1}{1-q} \log {\rm tr} (\rho_\mt{A}^{\,q})\,,
\labell{croak}
\end{eqnarray}
where $q$ is a real positive number called the order of the ERE. By design
given $S_q$ for a particular system, one can recover the usual EE with the
limit: $S_\mt{EE}=\lim_{q\to1} S_q$. However, the full family of ERE
clearly provides much more information about the density matrix, \ie in
principle, one can determine the full spectrum of eigenvalues of $\rho_\mt{A}$
\cite{calab8}.

Generally, it is also easier to calculate ERE than EE since the former avoids
the difficulty of calculating the logarithm of the density matrix. However, in
the context of quantum field theory, as we consider in this paper, the
calculation of ERE generally remains a challenging task. In
two-dimensional conformal field theory (CFT), there is a universal result for
the ERE of an interval of length $\ell$ \cite{cardy0}:
 \be
S_q(d=2)=\frac{c}{6} \left(1 + \frac{1}{q} \right)\log(\,\ell/ \delta)\,,
 \labell{twod}
 \ee
where $c$ is the central charge and $\delta$ is a short-distance regulator, in
the underlying CFT. In higher dimensions, the explicit results for ERE, as well
as our general understanding of these entanglement measures, are much more
limited \cite{cmt}. Essentially, any such results for ERE rely on the `replica
trick,' which requires evaluating the partition function on a $q$-fold cover of
the original background geometry \cite{cardy0,Callan:1994py}.

Recently, the AdS/CFT correspondence \cite{m1} has provided an alternative
perspective on EE for (certain) strong strongly coupled gauge theories. In
particular, the EE for a region in the boundary theory is calculated by
evaluating the black hole entropy formula on a corresponding extremal surface
in the dual bulk spacetime \cite{rt1}. At a fundamental level, this
prescription suggests that EE plays an important role in the quantum structure
of spacetime, \eg \cite{mvr,vm,arch}. An alternative approach to determine
holographic EE for spherical entangling surfaces was proposed in
\cite{Casini:2011kv} --- see also \cite{cthem} --- which has a number of
advantages. First, while the standard prescription \cite{rt1} only applies when
the bulk physics is described by Einstein gravity,\footnote{The original
prescription of \cite{rt1} was extended to holographic EE for Lovelock gravity
theories in \cite{GBent} --- see also the recent progress in \cite{aninda}.}
the approach of \cite{Casini:2011kv} can also be applied to bulk gravitational
theories which include arbitrary higher curvature interactions. Further, while
progress on a holographic ERE had been limited primarly to considering
two-dimensional boundary CFT's \cite{Headrick:2010zt},\footnote{See \cite{twod}
for more recent progress in this direction. Let us also add here that
\cite{aitor} made remarkable progress recently by extending the construction of
\cite{Casini:2011kv} to provide a derivation of the standard prescription for
holographic EE.} this alternative approach for spherical entangling surfaces is
easily generalized to calculating ERE for the same surfaces in higher
dimensional boundary CFT's \cite{Hung:2011nu}. Both of these features will be
essential for in the following where we use the holographic construction of
\cite{Casini:2011kv,Hung:2011nu} to study the ERE of four-dimensional ${\cal
{N}} = 4$ super-Yang-Mills (SYM) at strong coupling.

Let us briefly review the discussion of \cite{Casini:2011kv}. They considered
CFT's in flat $d$ dimensional Minkowski spacetime and computed the EE for a
spherical entangling surface of radius $R$. The causal domain of the region
enclosed by this sphere can be mapped using a conformal transformation to a
hyperbolic cylinder $R \times H^{d-1}$. The curvature scale of the hyperbolic
hyperplane is given by the radius of the original sphere. Further, if the CFT
began in the vacuum in flat space then, after the conformal mapping, one has a
thermal ensemble in the new geometry with temperature
\begin{equation}
T_0=\frac{1}{2 \pi R}\,.
\end{equation}
Hence, the EE for the spherical region of the CFT is equivalent to the thermal
entropy of the CFT at temperature $T_0$ on the hyperbolic cylinder. Further, if the theory admits an holographic dual, then,
using the AdS/CFT correspondence, we can translate this thermal entropy to the
horizon entropy of an appropriate black hole in the dual AdS spacetime. In
fact, the latter is found to be a so-called `topological black hole', \ie an
asymptotically AdS black hole for which the event horizon has the geometry
$H^{d-1}$. Indeed, this calculation does not rely on having Einstein gravity in the bulk, since we simply evaluate the horizon entropy using Wald's
formula \cite{Wald} in cases where there are higher curvature interactions.

As we mentioned previously, \cite{Hung:2011nu} showed that the above argument
can be extended to calculate holographic entanglement R\'enyi entropies for a
spherical entangling surfaces. First we observe that in the case of interest,
the trace of the power of the density matrix appearing in eq.~\reef{croak} can
be related to a thermal partition function with a new temperature, \ie
\begin{eqnarray}
tr(\rho_\mt{A}^{\,q})=\frac{Z(T_0/q)}{Z(T_0)^q}\,.
\end{eqnarray}
Then using standard thermodynamic relations, the ERE can be related to the
thermal free energy of the CFT on the cylinder with
\begin{eqnarray}
S_q=\frac{q}{1-q} \frac{1}{T_0} (F(T_0)-F(T_0/q))\,.
\labell{eqn-renyi-energy}
\end{eqnarray}
Alternatively, the ERE can be expressed in terms of the thermal entropy using,
\begin{eqnarray}
S_q=\frac{q}{q-1} \frac{1}{T_0} \int_{T_0/q}^{T_0}S_{therm}(T) \ dT\,.
\labell{eqn-renyi-entropy}
\end{eqnarray}
With these expressions in hand, we can evaluate the ERE in the holographic
framework by determining the required thermodynamic properties of the boundary
CFT on the hyperbolic cylinder by analyzing the same characteristics of the dual
family of topological black holes in the bulk theory.\footnote{In principle,
eqs.~\reef{eqn-renyi-energy} and \reef{eqn-renyi-entropy} can be used to
evaluate $S_q$ for any CFT without resorting to holography. Of course,
understanding the thermodynamic behaviour of the CFT on the hyperbolic cylinder
will be challenging but it has been successfully studied in certain cases
\cite{igor7}.}

As we said above, the aim of the present paper is to apply these expressions to
study the ERE for the most famous example of the AdS/CFT correspondence,
namely, four-dimensional ${\cal {N}} = 4$ $SU(N)$ super-Yang-Mills (SYM) dual to
type IIB string theory on an AdS$_5\times$S$^5$ background. This duality is
usually studied in the limit of an infinite number of colours and an infinite
't Hooft coupling, \ie $N\rightarrow\infty$ and $\lambda=g_{YM}^2 N\to\infty$
(while $\lambda/N\to0$), so that we get classical supergravity as the dual bulk
theory. In this limit, we only need to interpret the holographic calculations
found in \cite{Hung:2011nu} to apply them to the case of ${\cal N}= 4$ SYM.
However, we can also begin to relax the conditions on the coupling by analyzing
the effect of the leading stringy corrections to the low-energy effective
action. In type IIB string theory, the leading correction to the supergravity
action appears at $O(\alpha'^3)$ \cite{curv4,curv4a} and this naturally leads
to corrections of order $\lambda^{-3/2}$ in the dual SYM theory. In fact, by
examining the detailed form of these higher derivative corrections in the
effective action \cite{instant,Paulos:2008tn}, we can argue that they also
capture the first finite $N$ corrections proportional to $\lambda^{1/2}/N^2$
\cite{Myers:2008yi}.

Our results for the ERE will be very much analogous to holographic results for
the thermal entropy of SYM in the large $N$ limit. In particular, for a thermal
bath in flat space, the entropy density can be written as $s(T)=
\frac{2\pi^2}{3}\, N^2\, T^3\, f(\lambda)$ where the function $f(\lambda)$
encodes the dependence on the 't Hooft coupling. In the strong coupling limit,
holographic calculations indicate this function takes the form
\cite{Gubser:1998nz,Pawelczyk:1998pb}
\begin{eqnarray}
f(\lambda\to\infty)= \frac{3}{4} + \frac{45}{32}
\frac{\zeta(3)}{\lambda^{3/2}}+ \cdots,
\labell{eqn-f-lambda}
\end{eqnarray}
where the $\cdots$ indicate terms involving higher inverse powers of $\lambda$.
While at weak coupling, a two-loop calculation is required to show \cite{loopy}
 \beq
f(\lambda\to0)= 1- \frac{3}{4\pi^2} \lambda+ \cdots,
 \labell{weaker9}
 \eeq
where now $\cdots$ indicate contributions with higher powers of $\lambda$. Of
course, there is a mismatch in this two limits by a celebrated factor of $3/4$.
However, the fact that the leading corrections are positive at strong coupling
and negative at weak coupling suggest that $f(\lambda)$ is continuous function
that interpolates smoothly between these two limits.

The rest of the paper is organized as follows: In section \ref{twos}, we review the
results for the ERE of a spherical entangling surface in SYM, both at weak
coupling \cite{Fursaev:2012mp} and strong coupling \cite{Hung:2011nu}. In
section \ref{sec-corrections}, we will determine the effect of the leading
$\alpha'^3$ corrections in the effective action. We also analyze corrections
that this interaction produces for the scaling dimension of twist operators
that arise in calculations of the ERE. Section \ref{discuss} presents a brief
discussion of our results. Appendix \ref{appendix} provides an alternate
calculation of the corrections to the ERE induced by the leading $\alpha'^3$ terms in
the type IIB action, which gives a consistency check for our analysis in
section \ref{sec-corrections}.

\section{ERE at weak and strong coupling}
\label{twos}

In this section, we review the two known calculations for ERE of spherical
surfaces of radius $R$ in ${\cal {N}} = 4$ $SU(N)$ super-Yang-Mills. The first
one was obtained in the limit of zero 't Hooft coupling, \ie free fields. The
second one, obtained via holography, gives the value of ERE at infinitely
strong coupling and infinite $N$. Using the holographic dictionary, we will
show how these two solutions relate to each other. While for general $q$ the
ERE's at strong and weak coupling do not agree, the hope would be that there
exists some continuous function of the coupling that interpolates between the
two solutions.

\subsection{ERE at weak coupling}

At weak coupling, ERE for spherical entangling surfaces of radius $R$ were
calculated in \cite{Fursaev:2012mp}. Here we are considering the free field
limit of the SYM theory and so one only needs to sum the contributions for a gauge
field, four complex Weyl fermions and six massless (conformally coupled)
scalars, all in the adjoint representation. That computation yields, in the
large $N$ limit,
\begin{eqnarray}
S_q (R) \simeq N^2 \left( \frac{1}{2\,q} \frac{R^2}{\delta^2} -
\frac{1+q+7q^2+15 q^3}{24 q^3} \log \left(\frac{R}{\delta}\right)  \right),
\labell{weak-ERE}
\end{eqnarray}
where $\delta$ is a short-distance cut-off.\footnote{In order to be consistent
with the cut-off used at strong coupling in \citep{Hung:2011nu}, we set
$\delta=1/\Lambda$, where $\Lambda$ is the UV cut-off used in
\cite{Fursaev:2012mp}} The first contribution is the expected area law term
but the coefficient of this power law divergence is not universal. Hence, we
focus on the $\log$ term, which is expected to be universal, \ie independent of
any regularization scheme. Therefore, this contribution will be the interesting one
to compare with the strong coupling results.

As we will be comparing ERE for different values of $q$, it will be useful to
write
\begin{eqnarray}
S^{\log}_q (R) = - N^2 s(q) \log \left(\frac{R}{\delta}\right),
\end{eqnarray}
where we are defining
\begin{eqnarray}
s^{\text{weak}}(q)= \frac{1+q+7q^2+15 q^3}{24 q^3}.
\end{eqnarray}
Note that with $q=1$, for which the ERE reduces to the entanglement entropy, we
have $s^{\text{weak}}(q=1)=1$.

\subsection{ERE at strong coupling} \labell{stronger}

At strong coupling, we will follow \cite{Hung:2011nu} and briefly review how
the holographic result is calculated. This review will also set the stage for  our calculation on the corrections to the ERE in the next section.

The dual gravitational theory is simply five-dimensional Einstein gravity with
negative cosmological constant, whose action is
\begin{eqnarray}
I_{bulk}=\frac{1}{2 \lp^3} \int d^5x \sqrt{- g} \left( R + \frac{12}{L^2} \right),
\labell{eqn-action-zeroth}
\end{eqnarray}
where we defined $\lp^3=8 \pi G_N$. The topological black hole that satisfies
the corresponding equations of motion has a metric given by
\begin{eqnarray}
ds^2= - \left( \frac{r^2}{L^2} \left(1-\frac{\omega^4}{r^4} \right) - 1 \right)
\frac{L^2}{R^2} dt^2 + \frac{dr^2}{\frac{r^2}{L^2}\left(1-\frac{\omega^4}{r^4}
\right)-1} + r^2 d\Sigma^2_3 , \labell{eqn-metric}
\end{eqnarray}
where $d\Sigma^2_3$ is the line element for $H^3$, the hyperbolic plane in
three dimensions with unit curvature. Note that we chose to include a factor
$L^2/R^2$ in $g_{tt}$ term to ensure that the curvature scale of the boundary
metric is $R$. That is, the boundary CFT lives on the hyperbolic cylinder
$R\times H^3$ with metric,
\begin{eqnarray}
ds^{2}_{bndry}=-dt^2+R^2\, d\Sigma^2_3\,. \labell{boundmet}
\end{eqnarray}
We always have the freedom to adjust this constant factor as is convenient
since it simply corresponds to a constant rescaling of the time coordinate. The
event horizon in the above metric \reef{eqn-metric} is found by setting
$g_{tt}$ to zero. The latter then yields a relation between $\omega$ and the
position of the event horizon $r_h$,
\begin{eqnarray}
\omega^4 = r_h^4 - L^2 r_h^2\,.
\labell{eqn-omega}
\end{eqnarray}

Another important aspect of this geometry is that the horizon `area' is
proportional to the volume of the hyperbolic plane, $V_{\Sigma_3} = \int
d\Sigma$, which is divergent. Of course, the latter is closely related to the
UV divergences appearing in the R\'enyi entropies
\cite{Casini:2011kv,Hung:2011nu}. Hence, in order to get a finite result, one
only integrates up to some maximum radius defined by the asymptotic cut-off
surface in the AdS spacetime. This regulator is defined in terms of $\delta$,
the short distance cut-off in the boundary theory. Then one can expand the
volume $V_{\Sigma_3}$ in powers of $R / \delta$ and of particular interest for the
present calculation, we can identify the universal log term
\begin{eqnarray}
V_{\Sigma_3, univ}= - 2 \pi \log \left(\frac{R}{\delta}\right).
\labell{eqn-volume}
\end{eqnarray}

In order to calculate ERE with eq.~(\ref{eqn-renyi-entropy}), we need to know
both the temperature and the entropy of this solution. As usual, the
temperature is defined as the Hawking temperature and the entropy is given by
the horizon area,\footnote{Of course, in the next section where we consider the
effect of higher curvature interactions in the gravity action, we replace this
Bekenstein-Hawking expression with Wald's entropy \cite{Wald}.} \ie
$S=\frac{2\pi A}{\lp^3}$. For the above geometry \reef{eqn-metric}, we find
\begin{eqnarray}
T(x) & = & T_0 \left( 2x-\frac{1}{x} \right) \equiv T_1(x), \labell{eqn-temp} \\
S(x) & = & 2 \pi V_{\Sigma_3} \left( \frac{L}{\lp} \right)^3 x^3, \labell{eqn-entropy}
\end{eqnarray}
where we have introduced the variable $x\equiv r_h/L$ and as before,
$T_0=1/(2\pi R)$. We also defined the zero'th order temperature as $T_1(x)$, as
it will prove useful for our analysis in section \ref{sec-corrections}. In
terms of the variable $x$, eq.~(\ref{eqn-renyi-entropy}) becomes
\begin{eqnarray}
S_q  = \frac{q}{q-1} \frac{1}{T_0} \int_{x_q}^1 S_{therm}(x) \frac{dT}{dx} dx\,.
\labell{eq-renyi-q}
\end{eqnarray}
Here $x_q$ is defined as the value of $x$ such that the temperature is
$T=T_0/q$. That is,
 \be
x_q=\frac{1}{4q}\left( 1 + \sqrt{1+8q^2}\right)\,.
 \labell{xq}
 \ee
With these results, it is possible to evaluate eq.~(\ref{eq-renyi-q}) and
our holographic calculation of the ERE yields
\begin{eqnarray}
S_q = \frac{\pi q}{q-1} V_{\Sigma_3} \left( \frac{L}{\lp}
\right)^3 (2-x_q^2 (1+x_q^2))\,.
\labell{sqoo}
\end{eqnarray}
To make contact between this formula and the weak coupling result
\reef{weak-ERE}, we use the holographic dictionary to relate the ratio $L/\lp$
with SYM variables, \ie $L^3/\lp^3 = N^2/4\pi^2$. Then using
eq.~(\ref{eqn-volume}), we find at strong coupling (and large $N$)
\begin{eqnarray}
S^{\log}_q (R) = - N^2 s(q) \log \left(\frac{R}{\delta}\right)\,,
\labell{universe0}
\end{eqnarray}
with
\begin{eqnarray}
s^{\text{strong}} (q) & = & \frac{q}{q-1}  \frac{2- x_q^2 (1+x_q^2) }{2}
\nonumber\\
 &=& \frac{1+q}{64 q^3}\,\frac{(5\sqrt{1+8q^2}-3)(1+\sqrt{1+8q^2})^2}{3+\sqrt{1+8q^2}}\,.
 \labell{outside}
\end{eqnarray}
In the second line above, we have expressed $s^{\text{strong}} (q)$ in a
convenient form which makes evident that there is no singularity at $q=1$.

As already observed in \cite{Fursaev:2012mp}, the first thing to note is that
$s^{\text{strong}} (q)\ne s^{\text{weak}} (q)$. This should come as no surprise
since there is no reason to believe that this universal contribution to the ERE
should not depend on the coupling in general. However, an exception is the case
of $q=1$ for which the ERE becomes the entanglement entropy. Now in general,
the universal contribution to the EE is determined by the central charges of
the underlying (four-dimensional) CFT \cite{solo} and for a spherical
entangling surface, one has $S^{\log}_\mt{EE}(R) \simeq - 4a \log
\left(R/\delta\right)$ with $a=N^2/4$ for the ${\cal N}=4$ SYM theory (at large
$N$). Since the central charges are protected by supersymmetry, this result
must be independent of the `t Hooft coupling  --- \eg see \cite{stefan}.
Examining the above results for the ERE at $q=1$, we indeed find:
$s^{\text{strong}} (1)=1=s^{\text{weak}} (1)$. That is, with $q= 1$, we find
that universal contributions to the ERE agree at weak and strong coupling. In
figure \ref{fig 1}, we show the behaviour of the ratio between the strong and
weak coupling results for $s(q)$ as a function of $q$.
\begin{figure}
\setlength{\abovecaptionskip}{0 pt}
\centering
\includegraphics[scale=1.5]{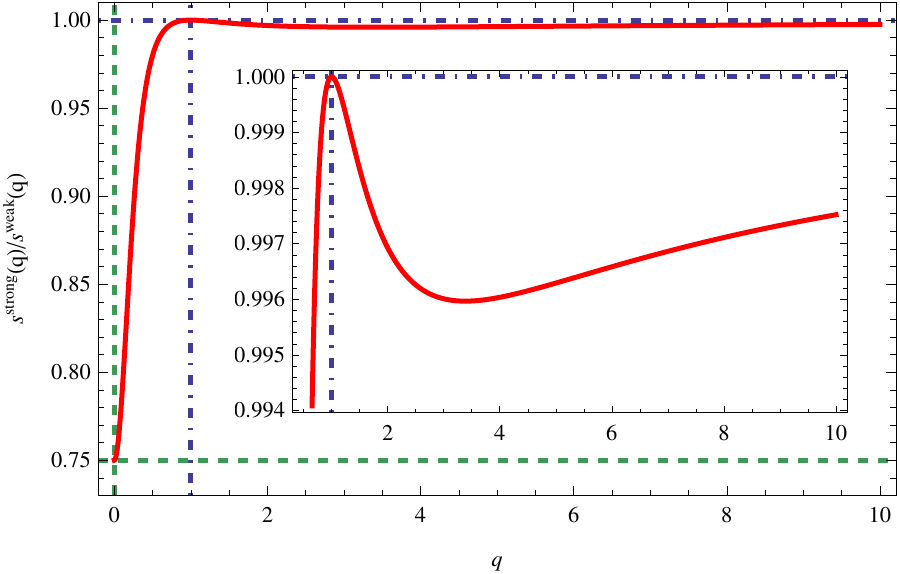}
\caption{Ratio between Entanglement R\'enyi Entropies at strong coupling over weak
coupling as a function of $q$. The green dashed line shows the limit of
$q\rightarrow0$, while the blue dashed-dotted line shows the result for
Entanglement Entropy ($q\rightarrow 1$). The inset zooms near the region of
$q\sim1$. } \labell{fig 1}
\end{figure}

As we see from figure \ref{fig 1}, another interesting limit is $q\rightarrow
0$.\footnote{Let us note that the limits, $q\to0$ and $\infty$, yield what are
known as the Hartley entropy and the min-entropy, respectively
\cite{Headrick:2010zt}. In particular, one finds $S_0 = \lim_{q\rightarrow 0}
S_q = \log ({\cal {D}})$ where ${\cal {D}}$ is the number of nonvanishing
eigenvalues of $\rho_\mt{A}$ and $S_\infty = \lim_{q\rightarrow \infty} S_q = -
\log (\lambda_1)$ where $\lambda_1$ is the largest eigenvalue. However, these
expressions involve the entire (regulated) ERE whereas our discussion here and
throughout the paper focuses only on the subleading universal contribution.}
The ratio between the ERE is exactly $3/4$ at this point. This factor is, of
course, reminiscent of the $3/4$ that can be found in eq.~(\ref{eqn-f-lambda})
for the ratio between the strong and weak coupling results for the thermal free
energy of ${\cal {N}}=4$ SYM. Examining the expression in
eq.~(\ref{eqn-renyi-energy}) more closely, one can see that in the limit of $q$
going to zero, the ERE must be proportional to the free energy of the theory in
the plane. That is, when $q\rightarrow0$, we have a divergent temperature
$T_0/q \rightarrow \infty$. As the temperature goes to infinity, the curvature
scale of the hyperbolic plane becomes negligible and then the free energy will
simply grow as the fourth power of the temperature, \ie as $q^{-4}$. Hence, in
this limit, the term $F(T_0/q)$ dominates the expression in
eq.~(\ref{eqn-renyi-energy}) and we find $S_q\simeq
-\frac{q}{T_0}F(T_0/q)=\frac{1}{4} S_{therm}(T_0/q)$ where both the free energy
and the thermal entropy are evaluated in flat space here. Hence we recover the
factor of $3/4$ observed previously in (flat space) thermal calculations for
the ${\cal {N}}=4$ SYM theory \cite{peet}. The discussion here provides a
specific example of the general result that the $q\to0$ limit of the ERE in any
CFT is governed by the high temperature behaviour of the theory
\cite{Swingle:2013hga}.

Finally, it is also interesting to note that in the limit of
$q\rightarrow\infty$, the ERE's at strong and weak coupling coincide. We have
no intuition for why the results agree in this low temperature limit and it appears
that it is simply a coincidence which does not survive after finite coupling
corrections are taken into account.

Knowing the results for the (universal contribution to the) ERE at both weak
and strong coupling, we are ready to start the study of higher order
corrections for the holographic calculations. From the study in this section, we should expect the ERE
corrections to take certain specific values, at least, for $q\rightarrow0$ and
for $ q\rightarrow1$. For the former, we expect this correction to be equal to
the correction to the free energy (as in eq.~(\ref{eqn-f-lambda})), while for
the latter we expect to find no correction.

\section{ERE corrections} \labell{sec-corrections}

We start this section by discussing the first corrections to the standard
supergravity action arising from the $\alpha'$ expansion in type IIB
superstring theory. It is known that the first corrections arise at the order
$\alpha'^3$, so that
\begin{eqnarray}
I_{IIB}= I_{IIB}^{(0)} + \alpha'^3\, I_{IIB}^{(1)} + \cdots,
\labell{eqn-typeIIB}
\end{eqnarray}
where the $\cdots$ indicate terms involving higher powers of $\alpha'$. Above,
$I_{IIB}^{(0)}$ indicates the usual ten-dimensional effective action of type
IIB supergravity \cite{john}. Early studies of string scattering amplitudes
\cite{curv4} and two-dimensional sigma-models \cite{curv4a} indicated the
appearance of quartic curvature interactions in $I_{IIB}^{(1)}$. It is also
important to consider contributions of the Ramond-Ramond five-form to this
action in order to properly determine the $\alpha'^3$ corrections for  phenomena
in the AdS$_5\times$ S$^5$ compactification. These were carefully analyzed in
\cite{Paulos:2008tn} and it was shown that in general, $I_{IIB}^{(1)}$ depends
on complicated combinations of the Weyl tensor and the Ramond-Ramond five-form
in ten dimensions. However, examining the equations of motion showed that the
higher order five-form terms all vanish for solutions of the form ${\cal
A}_5\times$ S$^5$, where ${\cal A}_5$ is a negatively curved Einstein manifold
\cite{Myers:2008yi}. In particular, the latter family of backgrounds would include the
topological black holes in AdS, which are of interest here. Moreover, in
\cite{Buchel:2008ae}, the dimensional reduction to five dimensions of the pure
curvature interaction in $I_{IIB}^{(1)}$ was analyzed, showing that it is
consistent to work only with a reduced action constructed from the
five-dimensional Weyl tensor. Hence for our holographic studies, it suffices to
work with the following five-dimensional action:
\begin{eqnarray}
I = I_{\text{bulk}} + \gamma \ I_{\text{Weyl}}\,,
\labell{fived}
\end{eqnarray}
where $I_{\text{bulk}}$ is the leading gravity action given in
eq.~(\ref{eqn-action-zeroth}). Further, $I_{\text{Weyl}}$ contains the
interaction quartic in the Weyl curvature,
\begin{eqnarray}
I_{\text{Weyl}}= \frac{L^6}{2 \lp^3} \int d^5x \sqrt{-g} \ \ W\,,
\labell{eqn-action-weyl}
\end{eqnarray}
with \cite{Gubser:1998nz,Buchel:2004di}
\begin{eqnarray}
W =  C^{hmnk} C_{pmnq} C_{h}^{\ rsp} C^{q}_{\ rsk}
 + \frac{1}{2}  C^{hkmn} C_{pqmn} C_h^{\ rsp} C^{q}_{\ rsk}\ ,
\labell{w}
\end{eqnarray}
where $C_{abcd}$ is the Weyl tensor in five dimensions
\begin{eqnarray}
C_{abcd}=R_{abcd} - \frac{2}{3} \left( g_{a[c} R_{d]b} - g_{b[c} R_{d]a} \right)
+ \frac{1}{6} R  g_{a[c} g_{d]b}\,.
\labell{eqn-c}
\end{eqnarray}
In eq.~\reef{fived}, the factor of $\alpha'^3$ has been absorbed into the
dimensionless coupling: $\gamma=\frac{1}{8} \zeta(3) \alpha'^3/L^6$. Hence
according to the standard AdS/CFT dictionary, this term in the gravity action
will yield corrections of order $\lambda^{-3/2}$ in the description of the dual
SYM theory. However, a more careful study of the origin of this higher
curvature interaction \cite{instant} allows one to argue that it also describes the
corrections at order $\lambda^{1/2}/N^2$ \cite{Myers:2008yi}
--- we return to this point in section \ref{discuss}.

Given the above action \reef{fived}, we proceed as follows: In the next
section, we calculate the $O(\gamma)$ corrections to the metric and Hawking
temperature of the topological black hole that appeared in the holographic
calculation of the ERE, as described in section \ref{stronger}. The ERE is
related to the thermal free energy on the hyperbolic cylinder with
eq.~\reef{eqn-renyi-energy}, which in turn is evaluated through the Euclidean
on-shell action in the bulk gravitational theory. We evaluate the latter in two
steps. First in section \ref{the-corrected-ERE}, we use the above perturbations
of the metric and temperature to determine the $O(\gamma)$ correction to
leading term in the action, $I_{\text{bulk}}$ (plus the corresponding boundary
terms). Then in section \ref{sec-simple}, we determine the corrections to the
ERE coming from $I_{\text{Weyl}}$. Since $I_{\text{Weyl}}$ already appears with
a pre-factor $\gamma$, we can evaluate this on-shell action with the original
black hole background. In the end, it turns out that this latter simpler
calculation yields the entire correction to the ERE. Finally in section
\ref{twist}, we follow the discussion of \cite{Hung:2011nu} to determine the
$O(\gamma)$ corrections to the scaling dimension of the twist operators
appearing in a standard calculation of the ERE.

\subsection{The perturbed metric} \labell{change}

In this subsection, we determine how the metric and the Hawking temperature are
modified by including the $I_{\text{Weyl}}$ term in the action \reef{fived}. As
this interaction only represents the first term in an infinite expansion, we
work perturbatively in $\gamma$ to find the corrections. We parametrize the
perturbed metric as
\begin{eqnarray}
ds^2 = & - & \left( \frac{r^2}{L^2} \left(1-\frac{\omega^4}{r^4} \right) - 1 \right) \left( 1 + \gamma f_2(r) \right) \frac{L^2}{R^2} dt^2 + \nonumber \\
&  & + \frac{dr^2}{\left(\frac{r^2}{L^2}\left(1-\frac{\omega^4}{r^4} \right)-1 \right) \left( 1 + \gamma f_3(r) \right)} + r^2 d\Sigma^2_3 ,
\labell{eqn-metric-perturbed}
\end{eqnarray}
where $f_2(r)$ and $f_3(r)$ are different functions that depend only on $r$.
This parametrization was chosen so that the coordinate position of the horizon
does not change, \ie $r_h$ is the same function of $\omega$ as in
eq.~(\ref{eqn-omega}). To find the equations of motion for $f_2$ and $f_3$, we
substitute the above metric into the action \reef{fived} and expand to second
order in $\gamma$ --- the first order variation vanishes identically, because
the leading order background solves the original equations of motion. The
resulting equations of motion for $f_2$ and $f_3$ are then
\begin{eqnarray}
f_3'(r) & + & \frac{2 r \left(L^2-2 r^2\right)}{L^2 r^2-r^4+\omega^4} \ f_3(r) + \nonumber \\
& + & \frac{20 \omega^{12} \left(160 L^2 r^2-144 r^4+171 \omega^4\right)}{r^{13}
\left(L^2 r^2-r^4+\omega^4\right)} + {\cal {O}}(\gamma) =0\,, \\
 f_2'(r)  & + &  \frac{2 r \left(L^2-2 r^2\right) }{L^2 r^2-r^4+\omega^4} f_3(r)+\frac{20
 \omega^{12} \left(16 L^2 r^2+27 \omega^4\right)}{r^{13} \left(L^2 r^2-r^4+\omega^4\right)}
  + {\cal {O}}(\gamma) =0 \,.
  \nonumber
\end{eqnarray}
Solving for $f_2(r)$ and $f_3(r)$, we find
\begin{eqnarray}
f_2(r) & = & \frac{5 \omega^{12} \left(16 L^2 r^2-24 r^4+9 \omega^4\right)}{r^{12}
\left(L^2 r^2-r^4+\omega^4\right)}+\frac{C_1}{L^2 r^2-r^4+\omega^4}+C_2\,, \nn\\
f_3(r) & = & \frac{5 \omega^{12} \left(64 L^2 r^2-72 r^4+57 \omega^4\right)}{r^{12}
\left(L^2 r^2-r^4+\omega^4\right)}+\frac{C_1}{L^2 r^2-r^4+\omega^4}\,,
\end{eqnarray}
where $C_1$ and $C_2$ are integration constants. Examining the asymptotic
metric, we see that $C_2$ simply produces a rescaling the time coordinate and
since we want the solution to be asymptotically conformal to the boundary
metric \reef{boundmet}, we set $C_2$ to zero. To fix $C_1$, we notice that
there is a potential divergence in $f_2(r_h)$ and $f_3(r_h)$ at the event
horizon, \ie at $r=r_h$ satisfying eq.~\reef{eqn-omega}. Then we choose $C_1$
to avoid these divergences,
\begin{eqnarray}
C_1 = \frac{5 L^4 }{x^4}\,\left(x^2-1\right)^3 \left(15 x^2-7\right)\,,
\end{eqnarray}
which is written in terms of the dimensionless parameter $x= r_h/L$. It is
noteworthy that with this single choice, we avoid the potential singularities
in both $f_2(r_h)$ and $f_3(r_h)$. Further, with this choice, we ensure that
the coordinate position of the horizon remains at $r=r_h$.

We will also need the corrected Hawking temperature for this black hole
geometry \reef{eqn-metric-perturbed}. It is calculated in the usual way and we
obtain to first order in $\gamma$
\begin{eqnarray}
T = T_1 (x) \left(1 + \frac{\gamma}{2} \left(f_2(r_h) + f_3 (r_h) \right) +
 {\cal {O}}(\gamma^2) \right),
\labell{eqn-temperature-order1}
\end{eqnarray}
where $T_1(x)$ is the zero'th order temperature given in eq.~(\ref{eqn-temp}).
Note that even though the coordinate position of the event horizon is not
changed by the perturbation, the temperature receives a correction. As the
holographic calculation of $S_q$ involves choosing the specific temperature
$T_0/q$, we have to find the correction to the coordinate position of the event
horizon that gives this fixed temperature at first order in $\gamma$. In terms
of $x$, we have
\begin{eqnarray}
\hat{x}_q = x_q - 10\,\gamma\, \frac{\left(x_q^2-1\right)^3
\left(1+3 x_q^2\right)}{x_q^6 \left(1+2 x_q^2\right)} + {\cal {O}}\left(\gamma^2\right)\, ,
\labell{eqn-xq-1}
\end{eqnarray}
where $x_q=\frac1{4q}\left( 1 + \sqrt{1+8q^2}\right)$ is the result determined
previously for the unperturbed solution.

\subsection{Corrections to ERE}
\labell{the-corrected-ERE}

Using eq.~(\ref{eqn-renyi-energy}), we calculate the corrected ERE by
evaluating the free energy at $T=T_0$ and $T_0/q$ (with $q$ any positive real
number). In turn, the free energy is determined by evaluating the on-shell bulk
action, which we do in two steps. First in this section, we use the above
perturbations of the metric and temperature to determine the $O(\gamma)$
correction coming from the leading term in the action \reef{fived} and we leave
the contribution from $I_{\text{Weyl}}$ to the next section. Of course,
evaluating $I_{\text{bulk}}$ on-shell by itself yields a divergent result,
which needs to be regularized. Holographic renormalization
\cite{Henningson:1998gx, Balasubramanian:1999re} provides a framework for the
latter, where boundary terms (that do not affect the equations of motion) are
added to the usual action to produce a finite result. In \cite{Emparan:1999pm},
it is shown that the full gravitational action (with Euclidean signature) can
be written in terms of three contributions:
\begin{eqnarray}
I_{\text{AdS}}= I_{\text{bulk}} (g_{ij})+I_{\text{surf}} (g_{ij})+I_{\text{ct}}
 (h_{ij})\,,
 \labell{totall}
\end{eqnarray}
where $I_{\text{bulk}}$ is the Einstein-Hilbert action \reef{fived},
$I_{\text{surf}}$ is the Gibbons-Hawking-York (GHY) term and $I_{\text{ct}}$ is
the counterterm action, that is only a function of the boundary metric
$h_{ij}$. If the bulk spacetime is five-dimensional with a four-dimensional
boundary, the GHY and counterterm actions are given by
\begin{eqnarray}
I_{\text{surf}} & = & - \frac{1}{\lp^3} \int_{\partial {\cal {M}}} d^4x\,
\sqrt{h}\, K  \    ,  \labell{boundact}\\
I_{\text{ct}} & = &  \frac{1}{\lp^3} \int_{\partial {\cal {M}}} d^4x\, \sqrt{h}\,
 \left[ \frac{3}{L} + \frac{L}{4} {\cal {R}} \right]\,,\nn
\end{eqnarray}
where $\partial {\cal {M}}$ corresponds to the boundary manifold with metric
$h_{ij}$. Further, $K = h^{ij} \nabla_i n_j$ is the trace of the extrinsic
curvature\footnote{Here, as in \cite{Emparan:1999pm}, the boundary metric is
defined as $h_{ij}=g_{ij}-n_i\,n_j$  with $n^i$ being an outward pointing unit
normal vector to the boundary $\partial {\cal {M}}$.} and ${\cal {R}}$ is the
Ricci scalar for the induced boundary metric. Then the usual procedure is to
evaluate the action with a cut-off surface at some large radius $R_{max}$. With
this geometric cut-off, each of the contributions in eq.~\reef{totall} is
finite but the potential divergences cancel between the bulk and surface
actions. Then taking the limit of $R_{max}\to\infty$ yields a finite result for
the on-shell action and hence for the free energy.

Of course, in the following, we carry out all of these calculations to first
order in $\gamma$. What we will find is that even though there are finite terms
at first order coming from each of the different terms in the full action
\reef{totall}, once we combine these and write the final result as a function of temperature (or
alternatively in terms of $q$), all of the $O(\gamma)$ contributions will be canceled. Hence,
the only $O(\gamma)$ contribution to the ERE will be that coming from
$I_{\text{Weyl}}$ in the next section. Here we might add that, as we will also
see in the next section, $I_{\text{Weyl}}$ yields a finite result when
evaluated on-shell and so no additional boundary terms are needed to regulate
it.

First we show how the divergences cancel between the three terms in the action
\reef{totall}. To first order in $\gamma$, evaluating $I_{\text{bulk}}$
on-shell yields
\begin{eqnarray}
I_{\text{bulk}}^{\text{on-shell}} & = & \beta \frac{V_{\Sigma_3}}{R} \left( \frac{L}{\lp} \right)^3
\Bigg(  -x^4+x_{\max }^4 +
\labell{eqn-bulk-action-onshell} \\
 & & \left. + \gamma \left[ \frac{15 \left(1-9 x^2\right) \left(x^2-1\right)^3}{2 x^4}+
 {{O}}\left(\frac{1}{x_{\max }}\right) \right] + {{O}} \left(\gamma^2\right) \right)\,,
 \nonumber
\end{eqnarray}
where $x_{max} \equiv R_{max} / L$. In the end, we take $x_{max}$ to infinity
and so we do not need to keep track of the ${{O}}\left(x_{\max }^{-1}\right)$
terms. Note that eq.~\reef{eqn-bulk-action-onshell} still contains a finite
term at order $\gamma$, which can be seen as coming from a boundary term at the
horizon. For the GHY term, we obtain
\begin{eqnarray}
I_{\text{surf}}^{\text{on-shell}} & = & \beta \frac{V_{\Sigma_3}}{R} \left( \frac{L}{\lp} \right)^3
 \Bigg( 2 x^2 \left(x^2-1\right)+3 x_{\max }^2-4 x_{\max }^4 + \nonumber \\
&  & \left. + \gamma \left[\frac{10  \left(15 x^2-7\right)\left(x^2-1\right)^3}{x^4}+{{O}}\left(\frac{1}{x_{\max }}\right)
 \right] + {{O}} \left(\gamma^2\right) \right)\,.
\end{eqnarray}
Again, this boundary contribution contains divergent terms at zero'th order,
\ie the $x_{max}^2$ and $x_{max}^4$ terms, but only finite terms at first order
in $\gamma$. Finally, the counterterm action yields
\begin{eqnarray}
I_{\text{ct}}^{\text{on-shell}} & = & \beta \frac{V_{\Sigma_3}}{R} \left( \frac{L}{\lp} \right)^3 \Bigg(
\frac{3}{8} \left(1+4 x^2-4 x^4\right)+3 x_{\max }^4-3 x_{\max }^2+
{{O}} \left(\frac{1}{x_{\max }}\right)  + \nonumber \\
 &  & \left. + \gamma \left[ \frac{15\left(15 x^2-7\right) \left(x^2-1\right)^3 }{2 x^4}+ {{O}}
 \left(\frac{1}{x_{\max }}\right)\right] + {{O}} \left(\gamma^2\right) \right)\,.
\end{eqnarray}
Note that this counterterm action not only cancels the divergences arising from
the bulk and surface actions but it also introduces an $x$-dependent term to
the action at zero'th order, {{\it i.e.} $O (\gamma^0)$. Again, there are no divergences but a finite
contribution appears at first order.  The fact that no divergences appear at
first order in any of these three actions is closely related to the fact that
both $f_2$ and $f_3$ decay rapidly with $r$ and so at the boundary, their
contributions are highly suppressed.

Combining all three contributions above and dividing by $\beta$, we find the
free energy $F=I/\beta$ is given by
\begin{eqnarray}
F(x) = \frac{V_{\Sigma_3}}{R} \left( \frac{L}{\lp} \right)^3 \left(
 \frac{3}{8} - \frac{x^2}{2} (1 + x^2) - \gamma\, \frac{10 \left(x^2-1\right)^3
\left(3 x^2+1\right)}{x^4}\right)\,.
\end{eqnarray}
Now this result can be used in eq.~(\ref{eqn-renyi-energy}) to determine the
contributions of $I_{\text{bulk}}$ to the ERE. Hence we recall that for $T=T_0$,
$x=1$ while for $T=T_0/q$, $x$ is defined to be $\hat{x}_q$ as given in
eq.~\reef{eqn-xq-1} to first order in $\gamma$. Combining these results, the
ERE becomes
\begin{eqnarray}
S_q = \frac{\pi q}{q-1} V_{\Sigma_3} \left( \frac{L}{\lp} \right)^3 & \Bigg[ & 2-\hat{x}_q^2 (1+\hat{x}_q^2)
+  \labell{eqn-renyi-xq}\\
 & - &  \gamma \frac{20 \left(x_q^2-1\right)^3 \left(3 x_q^2+1\right)}{x_q^4} + {{O}} \left( \gamma^2
  \right) \Bigg]\,,
 \nonumber
\end{eqnarray}
where we still do not incorporate the contribution from $I_{\text{Weyl}}$.  Of
course, the zero'th order term has precisely the expected form as given in
eq.~\reef{sqoo} from the previous calculations. However, note that this
expression is written in terms of $\hat{x}_q$ (rather than $x_q$), which
contains an $O(\gamma)$ term as shown in eq.~(\ref{eqn-xq-1}). Therefore a
first order correction is hidden in this `zero'th order' term, \ie
\begin{eqnarray}
\left. -\hat{x}_q^2 \left(1+\hat{x}_q^2\right) \right|_\gamma =
\gamma \frac{20 \left(x_q^2-1\right)^3 \left(3 x_q^2+1\right)}{x_q^4}.
\end{eqnarray}
Comparing the above expression with the explicit $O(\gamma)$ term in
eq.~\reef{eqn-renyi-xq}, we see that the two corrections precisely cancel.
Hence the final contribution here to the ERE is given by
\begin{eqnarray}
S_q = \frac{\pi q}{q-1} V_{\Sigma_3} \left( \frac{L}{\lp} \right)^3 &
 \Bigg[ & 2-x_q^2 (1+x_q^2) + {{O}} \left( \gamma^2 \right) \Bigg].
 \labell{eqn-renyi-xq-final}
\end{eqnarray}
The corresponding universal contribution to the ERE is given by precisely the
same expression as in eqs.~\reef{universe0} and \reef{outside}. However,
remember that here we have not yet taken into account the contribution from the
Weyl term \reef{eqn-action-weyl}, which will be considered in the next section.

\subsection{More corrections to ERE}
\labell{sec-simple}

In the previous section, we reproduced the contribution to ERE in
eq.~\reef{sqoo}, which was found with simply the Einstein action
\reef{eqn-action-zeroth}. However, the calculation there included the
$O(\gamma)$ corrections to the metric and Hawking temperature found in section
\ref{change} and there was a cancelation at this order which left $S_q$
unchanged. This result is incomplete though since we still have to account for
the contribution coming from $I_{\text{Weyl}}$. We calculate this correction in
the following.

Since this term already appears with a pre-factor $\gamma$ in
eq.~(\ref{fived}), it suffices to evaluate it on-shell the uncorrected metric
\reef{eqn-metric}. That is, any $O(\gamma)$ perturbation in the metric will
introduce a $\gamma^2$ term when included in $I_{\text{Weyl}}$. For the same
reason when evaluating the ERE with eq.~\reef{eqn-renyi-energy}, we can also
use $x_q$ as given in eq.~\reef{xq}, without the $O(\gamma)$ corrections
appearing in eq.~\reef{eqn-xq-1}.

Now, it is useful to note that $W$ can be written in the form
\cite{Buchel:2004di}
\begin{eqnarray}
W = B_{i j k l} \left( 2 B^{i k l j} - B^{l i j k} \right), \ \ \ \ \ \
B_{i j k l} = C^m{}_{i j n} \, C^n{}_{l k m}\,.
\labell{w-2}
\end{eqnarray}
Then, evaluating eq.~(\ref{w-2}) with the zero'th order metric
(\ref{eqn-metric}) yields
\begin{eqnarray}
W = \frac{180}{L^8} \frac{\omega^{16}}{r^{16}}\,.
\end{eqnarray}
This is quite similar to the result obtained for the planar AdS black hole
\cite{Gubser:1998nz}, with the difference that now $\omega$ is not $r_h$ but
instead, it is given by eq.~(\ref{eqn-omega}). Note that $W$ scales as
$r^{-16}$, so even with the $r^3$ coming from the determinant of the metric,
$I_{\text{Weyl}}$ will yield a finite result without any need of
regularization. After integrating then, we find
\begin{eqnarray}
I_{\text{Weyl}} = - \beta \frac{15}{2} \frac{V_{\Sigma_3}}{R} \left(\frac{L}{\lp}
 \right)^3  \frac{(1-x^2)^4}{x^4}\,.
\labell{eqn-weyl-onshell}
\end{eqnarray}

Now let us denote $\delta F=I_{\text{Weyl}}/\beta$, which takes the form
\begin{eqnarray}
\delta F= - \frac{15}{2} \frac{V_{\Sigma_3}}{R} \left(\frac{L}{\lp} \right)^3  \frac{(1-x^2)^4}{x^4}
\end{eqnarray}
To determine the corresponding contribution in eq.~(\ref{eqn-renyi-energy}), we
just evaluate this expression at $x=1$ and at $x=x_q$. It is interesting to
note that at $x=1$, \ie $r_h=L$, $I_{\text{Weyl}}$ vanishes. This vanishing can
be anticipated because when $T$ is exactly $T_0$, the topological black hole
solution is in fact just the pure AdS spacetime \cite{Casini:2011kv}, for which
the Weyl curvature vanishes. Evaluating the correction to ERE results in the
following universal contribution
\begin{eqnarray}
\delta S^{\log}_q(R) = - 8\gamma\, N^2\, s^{(1)}(q)\, \log \left(\frac{R}{\delta}\right) \,,
\labell{eqn-weyl-log}
\end{eqnarray}
with
\begin{eqnarray}
s^{(1)}(q) & = & -\frac{15}{16} \frac{q}{q-1}  \frac{(1-x_q^2)^4}{x_q^4} \nonumber\\
& = & 15\frac{(1+q)(1-q^2)^3}{q^3\,(3+\sqrt{1+8q^2})^4}\,.
\labell{eqn-s-weyl-log}
\end{eqnarray}
The final expression above was again chosen to make obvious that there is no
singularity at $q=1$.  Given the cancelation found in the previous section, the
above result is the entire $O(\gamma)$ correction to the ERE.

\begin{figure}
\setlength{\abovecaptionskip}{0 pt}
\centering
\includegraphics[scale=1.5]{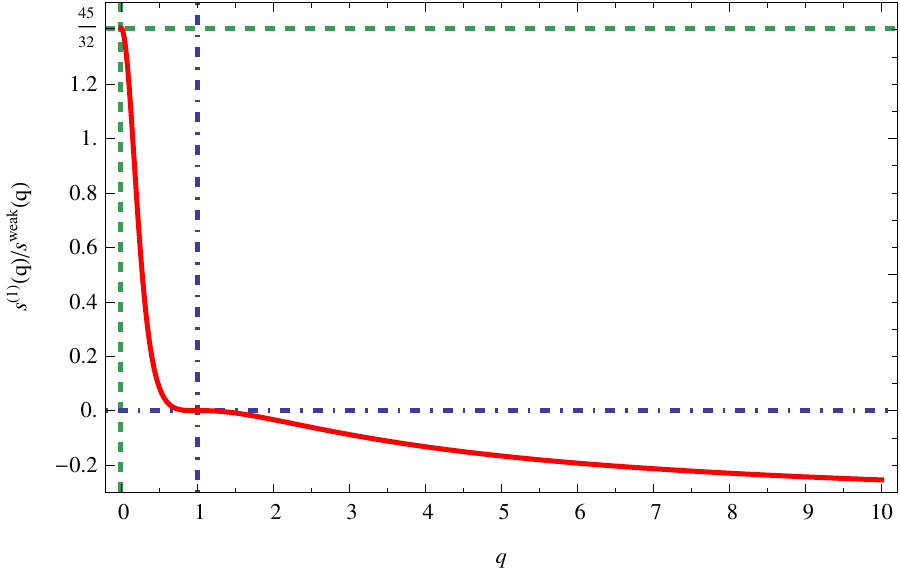}
\caption{Correction to the ERE due to $\alpha'^3$ effects. Again, the green dashed line
shows the limit of $q\rightarrow0$, while the blue dashed-dotted line shows the
result for entanglement entropy ($q= 1$).} \labell{fig 2}
\end{figure}
In figure \ref{fig 2}, we show $s^{(1)}(q)/s^{\text{weak}}(q)$ as a function of
$q$. We might note that the result for the corrections to ERE can be
anticipated in the two limits. First as explained in section \ref{stronger},
the ERE reduces to the entanglement entropy at $q=1$ and since the universal
contribution to the latter is determined by the central charge $a$, it must be
independent of the `t Hooft coupling. Hence the correction term
\reef{eqn-weyl-log} vanishes at $q=1$, as it is evident from the factor of
$(1-q^2)^3$ appearing in eq.~\reef{eqn-s-weyl-log}. The second case, where the
result can be anticipated is $q\to 0$, for which we find
$s^{(1)}(0)/s^{\text{weak}}(0)=45/32$. As also discussed in section
\ref{stronger}, the ERE is proportional to the flat space free energy in this
limit. Hence this fraction is precisely the pre-factor appearing in the
$O(\lambda^{-3/2})$ correction to the free energy appearing in
eq.~(\ref{eqn-f-lambda}).

Another interesting limit to consider is $q\rightarrow\infty$, for which
$s^{(1)}(\infty)/s^{\text{weak}}(\infty) = -3/8$. As was noted previously, even
though the leading terms at weak and at strong coupling coincide in this limit,
there is no obvious reason to believe that there should be no coupling
dependence. The fact that the correction \reef{eqn-weyl-log} is not vanishing
in this limit seems to indicate that the match found at leading order was
merely a coincidence.

\subsection{Scaling dimension of twist operators} \labell{twist}

Our holographic calculations of the ERE for a spherical entangling surface
take advantage of mapping the problem to a thermodynamic one, as described in
\cite{Casini:2011kv} --- see also \cite{cthem}. This approach contrasts with
`standard' field theoretic computations which make use of the replica trick,
\eg \cite{cardy0}. In this approach, one considers a path integral over $q$
replicas or copies of the original QFT and inserts {\it twist operators}
which open branch cuts between between these copies at the entangling surface.
In two dimensions, the twist operators are local conformal primary fields
\cite{cardy0} but in higher dimensions, they become ($d-2$)-dimensional
\emph{surface} operators. In practice the construction and properties of these
operators are not well understood for $d>2$.\footnote{However, see further
discussions in \cite{swingle,casini6}.} However, following \cite{Hung:2011nu},
we can use the thermal calculation for the ERE to determine their scaling
dimension $h_q$, in particular for holographic CFT's.

The scaling dimension $h_q$ for twist operators $\sigma_q$ in a
four-dimensional CFT is defined as follows: in flat Euclidean space, consider a
{\it planar} twist operator positioned at $x^1=0=x^2$ while it extends
throughout the remaining $x^3$ and $x^4$ directions. Now make an insertion of
the stress tensor operator at $x^\mu=\{y^1,y^2,y^3,y^4\}$ and define the
orthogonal distance between the two operators as $r^2= (y^1)^2 +(y^2)^2$. Then
the leading singularity in the corresponding correlator takes the following
form
 \bea
\langle T_{a b}\, \sigma_q \rangle&=& -\frac{h_q}{2\pi} \,\frac{\delta_{a
b}}{r^4} \,,\qquad \langle T_{a i}\, \sigma_q
\rangle=0\,,\nonumber \\
\langle T_{ij}\, \sigma_q \rangle&=& \frac{h_q}{2\pi} \,\frac{3\,\delta_{i
j}-4\,n_i\,n_j}{r^4}\,.
 \labell{weight}
 \eea
where $i,j=\lbrace 1,2\rbrace$, $a,b=\lbrace 3,4\rbrace$ and  $n^i=y^i/r$ is
the unit vector directed orthogonally from the twist operator to the
$T_{\mu\nu}$ insertion. Note that the form of this result is completely fixed
by translation and rotation symmetries, as well as $\langle T^\mu{}_{\mu}\,
\sigma_q \rangle=0=\nabla^\mu \langle T_{\mu\nu}\, \sigma_q \rangle$. Further
note that the correlators in eq.~\reef{weight} are implicitly normalized by
dividing out by $\langle \sigma_q \rangle$ but we left this normalization
implicit to avoid the clutter. Finally, let us add that we assume that
$T_{\mu\nu}$ corresponds to the total stress tensor for the entire $q$-fold
replicated CFT, \ie $T_{\mu\nu}$ is inserted on all $q$ sheets of the universal
cover.

Now as described in \cite{Hung:2011nu}, the conformal mapping between the
thermal state at temperature $T_0/q$ on the hyperbolic cylinder and the
$q$-fold cover of $R^4$ allows us to describe the leading singularity in
eq.~\reef{weight} and in particular, the scaling weight in terms of the thermal
properties of the CFT. The final result for $d=4$ is \cite{Hung:2011nu}
\begin{eqnarray}
h_q = \frac{2 \pi}{3} \ q \  R^4  \left( {\cal {E}} (T_0) - {\cal {E}} (T_0/q) \right)\, ,
\end{eqnarray}
where ${\cal {E}}= E/(R^3 V_{\Sigma_3})$ is the energy density of the thermal state on
$R\times H^3$. Our holographic calculations in the preceding sections have
determined the free energy and the temperature for SYM on the hyperbolic
cylinder and so it is straightforward to evaluate the scaling dimension of the
twist operators, including the leading corrections, using the standard
thermodynamic relation, $E= F - T \
\partial F/ \partial T$. The final result takes the form
\begin{eqnarray}
h_q & = & \pi \left(\frac{L}{\lp} \right)^3 q \ x_q^2 \left(1-x_q^2\right)  \nonumber \\
& & \qquad\times \left(1 + 5\gamma\, \frac{\left(1-x_q^2\right)^2 \left(-3+5 x_q^2+6 x_q^4\right)
}{x_q^6 \left(1+2 x_q^2\right)}+ \cdots \right) \\
& = & - \frac{\pi}{q^3} \left(\frac{L}{\lp} \right)^3 \Bigg(\frac{1}{32}\,
\frac{(1+\sqrt{1+8q^2})^3}{3+\sqrt{1+8q^2}}(1-q^2) +  \nonumber \\
& & \qquad\quad+\   10\, \gamma\, \frac{8q^2-3+9\sqrt{1+8q^2}}{\sqrt{1+8q^2}\,(3+\sqrt{1+8q^2})^3}(1-q^2)^3
+ \cdots \Bigg)\,.
 \labell{polyz}
\end{eqnarray}
The expression in the final line was chosen to make manifest the zero at $q=1$. The leading order term above applies for any four-dimensional CFT dual to
Einstein gravity in the bulk and is the same as the result reported in
\cite{Hung:2011nu}.

Examining the leading order result for $h_q$ for a broad class of holographic
theories in \cite{Hung:2011nu}, it was observed that with $d=4$
 \be
\partial_q h_q|_{q=1} = \frac{2 }{3 \pi}\, c\,.
 \labell{interest2}
 \ee
Recall that for ${\cal N}=4$ SYM with a $SU(N)$ gauge group, the two central
charges are given by $c=a=N^2/4$, in the large $N$ limit. From eq.~\reef{polyz}, we observe that the
$O(\gamma)$ correction to $\partial_q h_q|_{q=1}$ vanishes, providing evidence
that the expression in eq.~\reef{interest2} is independent of the coupling. In
fact, in \cite{twistop}, it was shown that eq.~\reef{interest2} is a general
formula which holds for any CFT. That is, this result does not rely on strong
coupling, large $N$ or even holography. Therefore since the central charges are
protected by supersymmetry in SYM, eq.~\reef{interest2} can not receive any
$\lambda$ dependent corrections and so the $O(\gamma)$ correction from
eq.~\reef{polyz} was required to vanish in evaluating $\partial_q h_q|_{q=1}$.
As is evident from eq.~\reef{polyz}, \ie from the factor of $(1-q^2)^3$, the
$O(\gamma)$ correction also vanishes for $\partial^2_q h_q|_{q=1}$ but is
nonvanishing for higher derivatives of the scaling weight. This observation can
again be related to further results found in \cite{twistop}. There, the conformal
mapping between the ERE for a spherical entangling surface and the thermal
state on the hyperbolic cylinder was used to show that $\partial^n_q
h_q|_{q=1}$ can be related to $n$- and $(n+1)$-point correlation functions of
the stress tensor. Hence $\partial^2_q h_q|_{q=1}$ is determined by the two-
and three-point function of the stress tensor. In the SYM theory, the
parameters controlling these correlators are still protected by supersymmetry
and so the $O(\gamma)$ contribution from eq.~\reef{polyz} is required to
vanish. In contrast, the higher derivatives of $h_q$ receive $O(\gamma)$
corrections, in accord with the expectation that the corresponding higher point
functions of the stress tensor will depend on the `t Hooft coupling.

\section{Discussion and conclusions} \labell{discuss}

In this paper, we have used holography to examine the entanglement R\'enyi
entropies for a spherical entangling surface in the ${\cal {N}}=4$ SYM theory
at strong coupling and large $N$. In particular, we have determined the leading
finite $\lambda$ (and finite $N$) corrections arising from higher curvature
corrections in the effective type IIB gravity action. Our results in the main
text were expressed in terms of a dimensionless expansion parameter
$\gamma=\frac{1}{8} \zeta(3)\, \alpha'^3/L^6$, which the standard AdS/CFT
dictionary translates to  $\gamma=\frac{1}{8} \zeta(3)\, \lambda^{-3/2}$ in the
dual SYM theory. However, a careful examination of the origin of the higher
curvature interaction \reef{eqn-action-weyl} reveals that the pre-factor
actually involves a modular form \cite{instant,Paulos:2008tn}. The latter
captures the behaviour for all values of the string coupling but remarkably
contains only string tree-level and one-loop contributions at weak coupling, as
well as an infinite series of instanton corrections. The tree-level term
corresponds to the $O(\lambda^{-3/2})$ correction but the one-loop term yields
a new correction at order $O(\lambda^{1/2}/N^2)$ \cite{Myers:2008yi}. Hence our
calculations also capture the leading finite $N$ corrections which appear at
this order. Note that this additional correction is enhanced by a factor of
$\lambda^{1/2}$ over what the naive large $N$ counting suggests would appear as
the first corrections. Combining the results in the main text then, we find
that the universal contribution to the ERE for a spherical entangling surface
in the ${\cal {N}}=4$ SYM theory at strong coupling and large $N$, behaves as
\begin{eqnarray}
S^{\log}_q (R) = - N^2  \log \left(\frac{R}{\delta}\right)\,\left[s^{(0)}(q)
+\left(\frac{\zeta(3)}{\lambda^{3/2}}+  \frac{\lambda^{1/2}}{48 N^2} \right)
 s^{(1)}(q)+\cdots\right]
\labell{universe1}
\end{eqnarray}
where
\begin{eqnarray}
s^{(0)} (q)
 &=& \frac{1+q}{64 q^3}\,\frac{(5\sqrt{1+8q^2}-3)(1+\sqrt{1+8q^2})^2}{3+\sqrt{1+8q^2}}\,,
 \labell{outside1}\\
s^{(1)}(q) & = & 15\frac{(1+q)(1-q^2)^3}{q^3\,(3+\sqrt{1+8q^2})^4}\,. \nonumber
\end{eqnarray}

One interesting observation in section \ref{sec-corrections} was that when the
metric with first order corrections was used to evaluate the contribution to
the ERE coming from the on-shell $I_{\text{bulk}}$, there was a cancelation at
$O(\gamma)$ and our final result took the same form \reef{sqoo} as without
considering the corrections. A similar result was found in \cite{Gubser:1998nz}
when calculating the leading corrections to the free energy for the planar AdS
black hole. In fact, general arguments indicate this behaviour will always occur
in a perturbative calculation of the type which we considered. Imagine
evaluating the on-shell action for a perturbed metric $(g_0+\gamma\,
g_1)_{ij}$, as follows
\begin{eqnarray}
I(g_0+\gamma g_1) & = & I_0 (g_0+\gamma g_1) + \gamma I_1 (g_0+\gamma g_1) + \cdots \nonumber \\
& = & I_0 (g_0) + \left. \frac{\delta I_0}{\delta g_{ij} } \right|_{g_0} \gamma\,
(g_1)_{ij} + \gamma I_1(g_0) + \cdots \nonumber \\
& = & I_0 (g_0) + \gamma I_1 (g_0) + \cdots\,,\labell{colon}
\end{eqnarray}
where $\cdots$ indicate higher order terms in the expansion. Between the second
and third lines we used that the variation of $I_0$ vanishes when evaluated on
a solution to the zero'th order equations of motion. However, there is some
subtetly in this argument, because the variation is actually only zero up to
total derivative terms. In fact, one can show that the extra terms appearing at
the leading order, \eg in eq.~(\ref{eqn-bulk-action-onshell}) for the bulk
action, are total derivatives. So why then do we find the cancelation at
$O(\gamma)$ in the final result. First, the profile of the perturbation decays
rapidly and so the boundary contribution produced by the total derivative
vanishes at asymptotic infinity. The other `boundary' to consider at the
horizon but in fact, the temperature is chosen to ensure that the Euclidean
geometry is smooth there, \ie there is no boundary at $r=r_h$. The appearance
of apparent contributions at this internal boundary arises because our
perturbative construction does not enforce this smoothness at every step.
However, as shown in section \ref{the-corrected-ERE}, when all of the
intermediate results are combined, the extra ``total derivative" terms cancel
in the final result and we are only left with $I_1 (g_0) =I_{\text{Weyl}}
(g_0)$ giving the full correction, as in eq.~\reef{colon}. It seems that it is
important that this simple general argument of expanding the action should only
be applied to the case of the renormalized (\ie the finite) action. We note again
that the same thing happens in \cite{Gubser:1998nz}. The bulk action appears to
receive corrections from the perturbed metric but after properly regularizing
and combining with the surface terms, the final result is independent of the
metric perturbation.

A consistency check of our results is given in appendix \ref{appendix}, where
we calculate ERE using a horizon entropy approach. The latter proceeds by
calculating the corrected horizon entropy, using Wald's formula \cite{Wald},
and Hawking temperature and then using eq.~(\ref{eqn-renyi-entropy}) to
determine the ERE. We show that this approach yields precisely the same results
as found in the main text. The advantage of this method is that one does not
need to consider surface terms and the holographic renormalization of the
action in order to get finite results. However, one still needs to calculate
metric perturbations as in section \ref{change} to get the temperature to first
order in $\gamma$ and then integrate eq.~(\ref{eqn-renyi-entropy}) to get ERE.

The latter also provides some insight on the behaviour of the corrections to
the ERE in eq.~\reef{universe1}. In particular, we note that $s^{(1)}(q)$
vanishes the limit $q\to1$ and actually, it has a cubic zero at this point, as
shown in eq.~\reef{outside1}. Now in the horizon entropy approach, the
correction to the ERE is determined by the correction to the horizon entropy
arising from $I_{\text{Weyl}}$ and as we showed in appendix \ref{appendix}, the
latter is cubic in the Weyl tensor. Further, the result at $q=1$ corresponds
precisely to the horizon entropy of the topological black hole with $T=T_0$,
but the latter background is, in fact, precisely the AdS vacuum solution for
which $C_{abcd}=0$! Hence this vanishing of the Weyl curvature and the cubic
dependence of the Wald entropy on $C_{abcd}$ can be seen as the gravitational
origin of the behaviour of $s^{(1)}(q)$ around $q=1$, which we noted above.

In part, our study was motivated by the fact that ERE were calculated for SYM
previously at both strong coupling \cite{Hung:2011nu} and weak coupling
\cite{Fursaev:2012mp}, but the results did not match in general. We reviewed
these previous calculations and compared the results in section \ref{twos}. Of
course, for such a comparison, we must focus on the universal log term in the
ERE and implicitly, we will be referring to this contribution throughout the
following.

With $q=1$, the ERE reduces to the entanglement entropy and the leading results
at strong and weak coupling were shown to agree in section \ref{twos}. As noted
there, this agreement is expected since in this case, the universal coefficient
is determined by the central charge $a$, which is protected by supersymmetry in
SYM. Of course then, this agreement at $q=1$ persists beyond the leading order
for the final result \reef{universe1}, which also includes the finite $\lambda$
and finite $N$ corrections. These terms have the potential to introduce
dependence of the EE on the 't Hooft coupling but as is evident from
eq.~\reef{outside1}, $s^{(1)}(q=1)=0$ and there are no higher order corrections
to the EE.

Another case where the leading results in section \ref{twos} were found to
agree in the strong and weak coupling limits was with $q\rightarrow \infty$.
However, there was no obvious reason to believe that the universal coefficient
in the ERE should be independent of the coupling in this limit. Our calculation
of the higher order corrections confirmed that this match at leading order was
only a coincidence. In particular, we see that $s^{(1)}(q \to \infty)  =
-15/64$ from eq.~\reef{outside1} and so there is an explicit $\lambda$
dependence in this limit for the ERE given in eq.~\reef{universe1}.

The final case where an interesting analytic comparison could be made between
strong and weak coupling was in the limit $q\to 0$. In this limit, the ERE is
probing the high temperature behaviour of the SYM theory. As seen with
$T_0/q\to \infty$ in eqs.~\reef{eqn-renyi-energy} and \reef{eqn-renyi-entropy},
the dominant contribution becomes $S_q\simeq -\frac{q}{T_0}F(T_0/q)=\frac{1}{4}
S_{therm}(T_0/q)$ where both the free energy and the thermal entropy can be
evaluated in flat space since the curvature scale of the hyperbolic plane
becomes negligible in this limit. Again we note that this is a specific example
of the general result recently discussed in \cite{Swingle:2013hga}. Comparing
the strong and weak coupling results in section \ref{twos}, we find
$s^{\text{strong}} (0)/ s^{\text{weak}} (0)=3/4$, which is precisely the famous
factor of $3/4$ found in comparing the thermal entropy at strong and weak
coupling for ${\cal {N}}=4$ SYM \cite{peet}. This matching also extends to the
higher order corrections found in section \ref{sec-corrections}. There we found
$s^{(1)}(0)/s^{\text{weak}}(0)=45/32$, which is precisely the prefactor found
for the $O(\lambda^{-3/2})$ correction to the thermal entropy in
eq.(\ref{eqn-f-lambda}). In fact, at weak coupling, we should also see the same
leading correction as in eq.~\reef{weaker9} appearing as the first correction
for $s^{\text{weak}}(q\to0)$ at $O(\lambda)$.

Examining the comparison of the leading results for strong and weak coupling
shown in figure \ref{fig 1}, we see that in general $s^{\text{strong}} (q)$
tends to be smaller than $s^{\text{weak}} (q)$. As discussed above, the
exceptions are $q=1$ and $q\to\infty$ where the two universal coefficients
coincide, but it is interesting to note that for no value of $q$ do we ever
find $s^{\text{strong}} (q)>s^{\text{weak}} (q)$. Further, while both of these
coefficients are positive throughout the full range of $q$, in figure {fig 2}
or eq.~\reef{outside1}, we see that $s^{(1)} (q)=s^{\text{Weyl}} (q)$, the
coefficient of the higher order corrections at strong coupling, changes sign at
$q=1$. In particular, for $q<1$, the correction to ERE is positive and so one
can imagine that as a function of the `t Hooft coupling, it is rising from the
strong coupling result towards the weak coupling answer, as depicted in figure
\ref{fig-q}a. This is similar to the situation for the thermal entropy,
described around eqs.~(\ref{eqn-f-lambda}--\ref{weaker9}). As discussed above,
at precisely $q=1$, the universal coefficient is independent of the coupling
and so the higher order corrections to the ERE vanish. Of course, we also have
$s^{(1)}(q)<0$ for $q>1$. This means that in this range of $q$, the first
corrections at the strong coupling are actually taking the ERE farther away
from the weak coupling limit, as shown in figure \ref{fig-q}c. While this does
not produce any inconsistency, the situation is perhaps slightly unusual. It
would be interesting to determine the corrections given by perturbation theory
for small coupling, or even the next term coming from the holographic
calculations at strong coupling, to have a more concrete idea of how ERE
behaves as a function of $\lambda$. In particular, in figure \ref{fig-q}, we
have assumed that ERE is a smooth function interpolating between the strong and
weak coupling limits, but it would interesting if instead there was a phase
transition at some intermediate value of the coupling.
\begin{figure}
\centering
\subfigure[$q<1$]{
\includegraphics[scale=0.5]{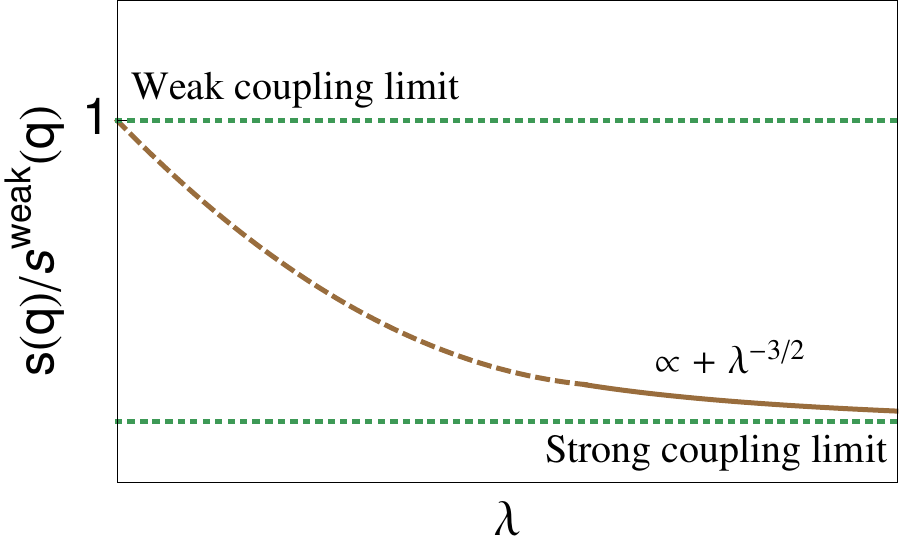}
\labell{fig-small-q} }
\subfigure[$q=1$ ]{
\includegraphics[scale=0.5]{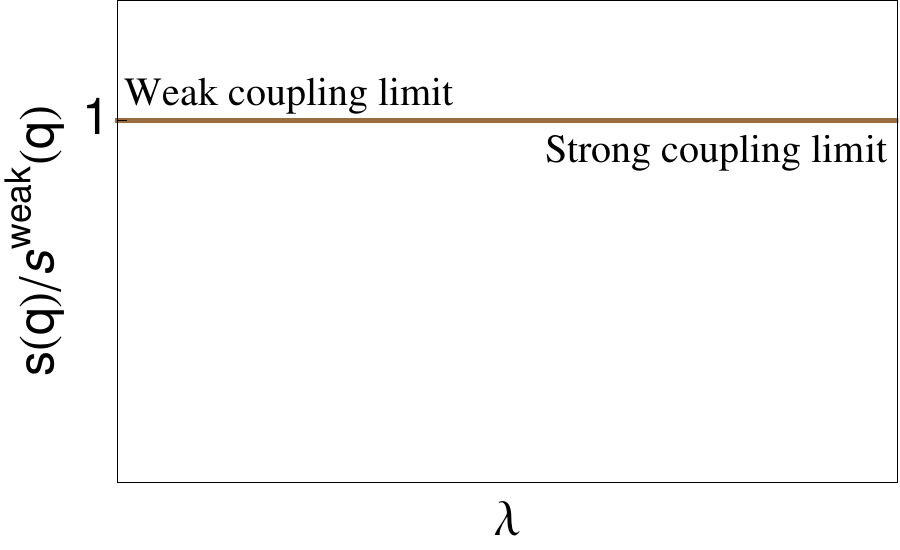}
\labell{fig-q-one} }
\subfigure[$q>1$]{
\includegraphics[scale=0.5]{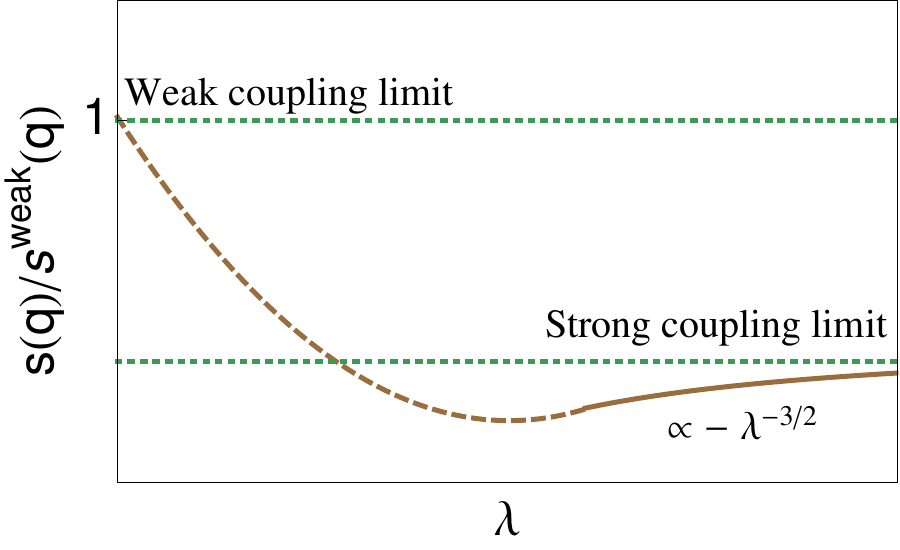}
\labell{fig-q-big} }
\caption{Schematic behaviour of ERE as a function of 't Hooft coupling for
different values of $q$. In each plot $q$ is fixed. For $q<1$, the first order
correction at strong coupling is positive; for $q=1$, there is no correction as
the EE is independent of $\lambda$; for $q>1$, the correction is negative.} \labell{fig-q}
\end{figure}

Lastly, in section \ref{twist}, we also calculated the the scaling dimension of
twist operators involved in calculating the ERE. Our calculations followed the
approach given in \cite{Hung:2011nu}, using the thermal energy density on the
hyperbolic cylinder. Our holographic calculations reproduced the leading strong
coupling result found in \cite{Hung:2011nu}, but we also found the leading
finite $\lambda$ and finite $N$ corrections for the SYM theory. Expressed in
terms of the boundary theory parameters, the scaling weight \reef{polyz}
becomes
\begin{eqnarray}
h_q& = & -  \frac{N^2}{4\pi q^3}  \Bigg(\frac{1}{32}\,
\frac{(1+\sqrt{1+8q^2})^3}{3+\sqrt{1+8q^2}}(1-q^2) +
 \labell{polyy}  \\
& & \qquad\quad+\   \frac{10}8\, \left(\frac{\zeta(3)}{\lambda^{3/2}}+
 \frac{\lambda^{1/2}}{48 N^2} \right)\, \frac{8q^2-3+9\sqrt{1+8q^2}}{\sqrt{1+8q^2}
 \,(3+\sqrt{1+8q^2})^3}(1-q^2)^3
+ \cdots \Bigg)\,.
\nonumber
\end{eqnarray}
Further following the discussion of \cite{Hung:2011nu,twistop}, it was observed
above that this result satisfies certain identities which are protected by
supersymmetry. For example, as given in eq.~\reef{interest2}, we have
$\partial_q h_q|_{q=1} = \frac{N^2}{6 \pi}$ and a similar identity will hold
for $\partial^2_q h_q|_{q=1}$. The coupling independence of these expressions
results from the factor of $(1-q^2)^3$ in higher order contributions in
eq.~\reef{polyy}. The higher derivatives of $h_q$ receive finite $\lambda$ (and
finite $N$) corrections because the coupling independence expectation does not
extend the corresponding higher point functions of the stress tensor. At
present, there are no comparable results at weak coupling with which we can
compare this scaling weight \reef{polyy} at strong coupling. We observe that
the approach for calculating $h_q$ in \cite{Hung:2011nu} does not rely on
holography and so could equally well be applied to SYM at weak coupling, \eg
using heat kernel techniques. This would, of course, be an interesting project
for future work.

\section*{Acknowledgements}
We thank Alex Buchel, Matt Headrick, Dmitri Fursaev and Brian Swingle for
useful discussions and acknowledge the use of Matt Headrick's
\textit{Mathematica} package
\href{http://people.brandeis.edu/~headrick/Mathematica/diffgeo.m}{diffgeo.m}.
Research at Perimeter Institute is supported by the Government of Canada
through Industry Canada and by the Province of Ontario through the Ministry of
Research $\&$ Innovation. The authors are also supported by an NSERC Discovery
grant. RCM also receives support from the Canadian Institute for Advanced
Research.

\appendix
\section{A horizon entropy approach to ERE}
\labell{appendix}

The aim of this Appendix is to provide an alternative way to calculate ERE. As
shown in eq.~(\ref{eqn-renyi-entropy}), one can obtain the ERE by knowing the
behaviour of both the thermal entropy and the temperature. Integrating by parts
in eq.~(\ref{eqn-renyi-entropy}) and writing the result in terms of variable
$x$, we find \cite{Hung:2011nu}
\begin{eqnarray}
S_q =  \frac{q}{q-1} \frac{1}{T_0} \left( \left. S(x) T(x) \right|^1_{x_q}
-\int_{x_q}^{1} \frac{dS}{dx} T(x) dx \right).
\end{eqnarray}

Of course, the thermal entropy corresponds to the horizon entropy in the dual
gravity theory. For theories with higher curvature interactions, one must
evaluate the Wald entropy \cite{Wald}
\begin{eqnarray}
S = -2 \pi \int_{\text{horizon}} d^3x \ \sqrt{h} \ \frac{\partial {\cal {L}}}{
\partial R_{abcd}} \ \hat{\varepsilon}_{ab}\, \hat{\varepsilon}_{cd}\,,
\labell{froggy}
\end{eqnarray}
where ${\cal {L}}$ is the Lagrangian for the particular gravitational theory
under consideration and $\hat{\varepsilon}_{ab}$ is the volume form in the
two-dimensional space transverse to the bifurcation surface of horizon. The
latter is normalized so that $\hat{\varepsilon}_{ab}\hat{\varepsilon}^{ab}=-2$.
Of course, for Einstein gravity with ${\cal {L}}=R/(2\lp^3)$, this expression
\reef{froggy} reduces to the Bekenstein-Hawking area law, \ie $S=2\pi {\cal
A}/\lp^3$. However, in the present case, we also need to analyze the
contribution coming from $I_{\text{Weyl}}$ in eq.~\reef{eqn-action-weyl}. To
incorporate this correction we first note that the geometry at the horizon is homogeneous,
so we may simplify eq.~\reef{froggy} as
\begin{eqnarray}
S =  -2 \pi  \ r_h^3 \ V_{\Sigma_3} \ \left.
 \frac{\partial {\cal {L}}}{\partial R_{abcd}} \
 \hat{\varepsilon}_{ab}\, \hat{\varepsilon}_{cd} \right|_{r=r_h}.
 \labell{tadpole}
\end{eqnarray}

Now given the five-dimensional action \reef{fived}, we have
\begin{eqnarray}
{\cal {L}} = \frac{1}{2 \lp^3} (R + \gamma W),
\end{eqnarray}
where we recall that $W$ is given by
\begin{eqnarray}
W & = &  C^{hmnk} C_{pmnq} C_{h}^{\ rsp} C^{q}_{\ rsk}
+ \frac{1}{2}  C^{hkmn} C_{pqmn} C_h^{\ rsp} C^{q}_{\ rsk}\ , \labell{weyl-appendix}
\end{eqnarray}
with
\begin{eqnarray}
C_{abcd} & = & R_{abcd} - \frac{1}{3} \left( g_{ac} R_{db} - g_{ad} R_{cb} -
g_{bc} R_{da} + g_{bd} R_{ca} \right) + 
\frac{1}{12} R  (g_{ac} g_{db} - g_{ad} g_{cb} )\,. \labell{riemann-appendix}
\end{eqnarray}
To compute the variation with respect to $R_{abcd}$ in eq.~\reef{tadpole}, it
is easiest to first vary with respect to the Weyl tensor and then with respect
to the Riemann tensor, \ie
\begin{eqnarray}
\frac{\partial {\cal {L}}_{\text{Weyl}}}{\partial R_{abcd}} = \frac{\partial
{\cal {L}}_{\text{Weyl}}}{\partial C_{efgh}} \cdot \frac{\partial C_{efgh}}{\partial
R_{abcd}}\,.
\end{eqnarray}
This is a rather tedious but straightforward calculation. Basically, each of
the $C$'s in $W$ will contribute with one term to the first variation above, while
each $R$, $R_{ab}$, $R_{abcd}$ will contribute with one term to the second one.
So in total we find 48 different terms. For example, the first term corresponds
to taking the derivative with respect to the first Weyl tensor in
eq.~(\ref{weyl-appendix}) and with respect to the first $R_{abcd}$ in
eq.~(\ref{riemann-appendix}) to produce
\begin{eqnarray}
\left( \left. \frac{\partial {\cal {L}}_{\text{Weyl}}}{\partial R_{abcd}}
\hat{\varepsilon}_{ab} \hat{\varepsilon}_{cd} \right|_{r=r_h} \right)_I =
\left. C_{pmnq} C_h^{\ rsp} C^q_{\ rsk} \hat{\varepsilon}^{hm} \hat{\varepsilon}^{nk}
\right|_{r=r_h} = -84 \ \frac{\omega^{12}}{r_h^{\ 12}}.
\end{eqnarray}
All the non vanishing terms yield an expression proportional to
${\omega^{12}}/{r_h^{\ 12}}$ when evaluated on the topological black hole solution and so one only has to sum all the different
contributions. In all, writing the result in terms of $x$ instead of $r_h$
yields
\begin{eqnarray}
S (x) =  2 \pi V_{\Sigma_3} \left( \frac{L^3}{\lp^3}\right)
x^3  \left(1 + 60 \gamma \frac{(x^2-1)^3}{x^6} \right).
\labell{eqn-entropy-app}
\end{eqnarray}

Notice that the zero'th order term is exactly the same as that in
eq.~(\ref{eqn-entropy}). To compare our previous results with
eq.~\reef{eqn-entropy-app}, it is simply to translate the free energy found
in the main text into the corresponding thermal entropy using
 \beq
S=-\frac{\partial F}{\partial T} = -
{\frac{\partial F}{\partial x}}\,\bigg/\,{\frac{
\partial T }{ \partial x}}\,.
 \labell{orange}
 \eeq
Having the expressions of both the free energy and the temperature to first
order in eqs.~(\ref{eqn-weyl-onshell}) and (\ref{eqn-temperature-order1}), we
can compute the thermal entropy to first order in $\gamma$ and we find that we
reproduce precisely eq.~(\ref{eqn-entropy-app}). Of course, if one gets the
same entropy in both cases, then the calculation for ERE must
also agree.

\end{document}